# Natural Deduction as Higher-Order Resolution
# (Revised version)

Lawrence C Paulson
Computer Laboratory
University of Cambridge

December 1985

**Abstract**

An interactive theorem prover, *Isabelle*, is under development. In LCF, each inference rule is represented by one function for forwards proof and another (a *tactic*) for backwards proof. In Isabelle, each inference rule is represented by a Horn clause. Resolution gives both forwards and backwards proof, supporting a large class of logics. Isabelle has been used to prove theorems in Martin-Löf's Constructive Type Theory.

Quantifiers pose several difficulties: substitution, bound variables, Skolemization. Isabelle's representation of logical syntax is the typed $\lambda$-calculus, requiring higher-order unification. It may have potential for logic programming. Depth-first subgoaling along inference rules constitutes a higher-order Prolog.

# Contents

# 1  Background

At least seven interactive theorem provers use the LCF framework. They differ primarily in what logic is used for conducting proofs. A new theorem prover, *Isabelle*, is intended to unify these diverging paths.

A recurring theme will be the relationship between *syntax* and *semantics*. We compute by means of syntax but think in terms of semantics. A system of inference rules is a syntactic codification of semantic concepts, and must be shown to respect them. Sets and functions are semantic; formal type theories and Zermelo-Fraenkel set theory are syntactic. The ultimate semantic notion is *truth*, faintly approximated by *theorems* of formal logic. Regarding an axiom system as Holy Writ blurs the distinction. Martin-Löf [22] discusses the evolution of formal logic from an intuitionistic viewpoint. There are many distinct semantic viewpoints, and accordingly many formal systems. A universal logic is too much to hope for. However, we all seem to write down formal proofs in the same way: by joining inference rules together.

Isabelle evolved by trial and error from implementations of Martin-Löf's Constructive Theory of Types. I started with ML and the LCF architecture, as described in the next section. Attempts to put *unification* into LCF pushed theorems from the dominant position, focusing attention on *inference rules*. Quantifiers posed an endless series of problems. Martin-Löf's syntactic theory of expressions, akin to the typed $\lambda$-calculus, gave a uniform framework for bound variables and substitution. This required higher-order unification, in general undecidable; the literature offered little practical advice. Quantifiers also introduced *dependencies* between variables in the proof, requiring some form of Skolemization.

As an example, the rules of Martin-Löf's Constructive Type Theory have been put into Isabelle, with appropriate tactics. The types of many constructions can be inferred automatically; simple functions can be derived interactively.

# 2  The LCF interactive theorem prover

Edinburgh LCF introduced a new approach to theorem proving: embed the formal logic in a programmable meta language, ML [15]. Terms and formulas are values in ML: they have an explicit tree structure and can be decomposed and built up by ML functions. Theorems are values of an abstract data type *thm*. In place of arbitrary constructor functions, inference rules map theorems to theorems. This is forwards proof. For backwards proof, goals, subgoals, and tactics are implemented on top of this abstract type. Each rule checks that it has received suitable premises, then generates the conclusion. Each tactic checks that it has received a suitable goal, then generates the subgoals giving the corresponding rule as validation. Milner [27] explains rules and tactics, while Gordon [12] works out the representation of a simple logic in ML.

Recent LCF proofs involve denotational semantics, verification of functional programs, and verification of digital circuits [30]. Logics include two versions of PP$\lambda$ (for domain theory) [15, 31], a Logic for Sequential Machines, two higher-order logics [8, 13], and two constructive type theories [9, 32]. Implementing a logic is a major undertaking: choosing a representation of formulas, implementing several dozen inference rules and tactics, implementing many more derived rules and higher level tools, implementing a theory database, writing a parser and printer, and documenting all these things.

## 2.1 Forwards proof and inference rules

*Inference rules* are functions from theorems to theorems; *axioms* are the built-in theorems of an LCF theory. Theorems can only be created by applying inference rules to axioms and other theorems. The logic of Cambridge LCF is PP$\lambda$, for reasoning in domain theory. The axiom of reflexivity for the partial ordering $\sqsubseteq$ is $\vdash \forall x.\, x \sqsubseteq x$, which is bound to the ML identifier *LESS_REFL* of type *thm*. The rule of conjunction introduction,

$$\frac{\Gamma \vdash A \quad \Delta \vdash B}{\Gamma, \Delta \vdash A \wedge B}$$

is bound to the ML identifier *CONJ* of type $thm \to thm \to thm$.

In natural deduction, each theorem is proved with respect to a set of assumptions. Conventional textbooks [38] treat the assumptions as leaves of the proof tree; when the assumption is discharged, the leaf is crossed out. LCF attaches the assumptions directly to the theorem, resulting in a sort of sequent calculus. The conjunction rule states that if the premises are $\vdash A$ and $\vdash B$ with assumptions $\Gamma$ and $\Delta$, then the conclusion is $\vdash A \wedge B$ under the union $\Gamma, \Delta$ of those assumptions.

The incorrect use of a rule is rejected using ML's exception mechanism: a function can *fail* instead of returning a result. The conjunction elimination rules are

$$\frac{\Gamma \vdash A \wedge B}{\Gamma \vdash A} \qquad \frac{\Gamma \vdash A \wedge B}{\Gamma \vdash B}$$

They are implemented by ML functions of type $thm \to thm$ that fail if the argument is not a conjunction. Modus Ponens, or implication elimination, is

$$\frac{\Gamma \vdash A \Rightarrow B \quad \Delta \vdash A}{\Gamma, \Delta \vdash B}$$

The function *MP* checks that its first argument has the form $\vdash A \Rightarrow B$ and that its second argument is $\vdash A$. If either condition is violated, *MP* fails.

Disjunction introduction raises a question. What if the conclusion contains formulas not in the premises?

$$\frac{\Gamma \vdash A}{\Gamma \vdash A \vee B}$$

In LCF, the formula $B$ is supplied as an extra argument. The ML function *DISJ1* has type $thm \to form \to thm$.

Implication introduction is also called the discharge rule. It discharges a formula from the set of assumptions, forming an implication:

$$\frac{\Gamma, A \vdash B}{\Gamma \vdash A \Rightarrow B}$$

The assumption $A$ is an extra argument; *DISCH* has type $form \to thm \to thm$. Notation: $\Gamma, A$ abbreviates $\Gamma \cup \{A\}$; the order of assumptions does not matter in classical first-order logic.

## 2.2 Quantifiers

Quantifiers bring in several concepts: substitution, bound variables, and parameters. A common convention for *substitution* writes the formula $A$ as $A[t]$ to emphasize that it may have free occurrences of $t$. Then $A[u]$ is the result after substituting $u$ for those occurrences.

In $\forall x.A[x]$, the variable $x$ is *bound*. The name is immaterial: this formula is equivalent to $\forall y.A[y]$. Naive substitution can change the meaning of a formula because of clashes among variable names. LCF renames bound variables when necessary; the Lisp code to do this is dismayingly complex.

A *parameter* stands for an arbitrary, fixed value in a proof,[1] as when we say, 'let $\epsilon$ be given.' LCF and many authors let any free variable serve as a parameter. Some authors regard parameters $a, b, \ldots$ as syntactically distinct from variables $x, y, \ldots$; parameters cannot be bound, and theorems contain no free variables. To be 'arbitrary', a parameter must not appear in certain formulas.

Consider the universal introduction rule. If the theorem $\vdash A[x]$ holds for arbitrary $x$, then $\vdash \forall y.A[y]$ holds. 'Arbitrary' means that the assumptions $\Gamma$ have no free occurrence of $x$:

$$\frac{\Gamma \vdash A[x]}{\Gamma \vdash \forall y.A[y]} \qquad x \text{ not free in } \Gamma$$

The restriction on the parameter $x$ prevents contradictions such as

$$\begin{array}{ll}
x = 0 \vdash x = 0 & \text{assumption rule} \\
x = 0 \vdash \forall y.y = 0 & \text{\emph{incorrect} use of } \forall\text{-introduction} \\
\vdash x = 0 \Rightarrow \forall y.y = 0 & \Rightarrow\text{-introduction} \\
\vdash \forall z.(z = 0 \Rightarrow \forall y.y = 0) & \forall\text{-introduction} \\
\vdash 0 = 0 \Rightarrow \forall y.y = 0 & \forall\text{-elimination with } 0 \\
\vdash \forall y.y = 0 & \Rightarrow\text{-elimination with } 0 = 0 \\
\vdash 1 = 0 & \forall\text{-elimination with } 1
\end{array}$$

In LCF, universal introduction is the function *GEN* of type *term* $\to$ *thm* $\to$ *thm*. Its first argument is the parameter $x$. If *GEN* finds an occurrence of $x$ in the assumptions, then it fails.

The existential introduction rule is ambiguous in forwards proof:

$$\frac{\Gamma \vdash A[t]}{\Gamma \vdash \exists x.A[x]}$$

Which occurrences of $t$ should be replaced by $x$? The premise $\vdash 0 = 0$ gives four different conclusions:

$$\vdash \exists x.\, x = x \qquad \vdash \exists x.\, 0 = x \qquad \vdash \exists x.\, x = 0 \qquad \vdash \exists x.\, 0 = 0$$

To resolve the ambiguity, the function *EXISTS* takes two extra arguments: the term $t$ and the formula stating the conclusion $\exists x.A[x]$.

## 2.3 Backwards proof and tactics

At the lowest level LCF executes forwards proof, but it also supports backwards proof. The *goal* $\Gamma \vdash^{?} B$ expresses the desired theorem $\Gamma \vdash B$. A *tactic* is a function that reduces a goal to a list of subgoals. The conjunction of the subgoals should imply the goal, so repeated use of tactics forms an AND tree of goals. In LCF, this top-down decomposition of the goal must be followed by a bottom-up reconstruction of the proof. Tactics provide high-level assistance in the search for a proof, but a theorem can only be produced by executing the primitive inference rules. So each tactic returns a *validation* function. The validation ought to produce a theorem achieving the goal, given theorems achieving the subgoals. If it does not, the tactic is *invalid*, and can only lead the user down an incorrect path. We cannot be certain of the validations until they are applied, usually at the very end of the proof.

---

[1]This paper uses *parameter* only to mean this: never to mean an argument to a function.

In sum, a tactic maps a goal $\Gamma \stackrel{?}{\vdash} B$ to the pair

$$([\Gamma_1 \stackrel{?}{\vdash} B_1, \ldots, \Gamma_n \stackrel{?}{\vdash} B_n], \textit{validation})$$

of subgoals and validation. Every tactic has the type

$$\textit{goal} \rightarrow ((\textit{goal list}) \times (\textit{thm list} \rightarrow \textit{thm})) \,.$$

For example, the tactic for conjunction introduction is called $CONJ\_TAC$. It maps any goal of the form $\Gamma \stackrel{?}{\vdash} A \wedge B$ to the two-element goal list $[\Gamma \stackrel{?}{\vdash} A, \Gamma \stackrel{?}{\vdash} B]$. The validation, which calls the rule $CONJ$, maps the two-element theorem list $[\Gamma \vdash A, \Gamma \vdash B]$ to the theorem $\Gamma \vdash A \wedge B$.

Constructing the validation is routine, but there is no uniform method because the rule may be a function taking extra arguments. The tactic for the discharge rule, $DISCH\_TAC$, maps a goal $\Gamma \stackrel{?}{\vdash} A \Rightarrow B$ to the one-element goal list $[\Gamma, A \stackrel{?}{\vdash} B]$. It adds the antecedent, $A$, to the assumptions in the subgoal. The validation calls the rule $DISCH$, passing the extra argument $A$. The universal introduction tactic, $GEN\_TAC$, maps a goal $\Gamma \stackrel{?}{\vdash} \forall x.A[x]$ to one subgoal $\Gamma \stackrel{?}{\vdash} A[x']$, choosing the parameter $x'$ to differ from all the variables in the assumptions. The validation calls $GEN$, passing it $x'$.

When a premise of a rule contains a term not in the conclusion, its tactic requires extra arguments to compute the subgoals. The tactic for existential introduction, $EXISTS\_TAC$, takes a term $t$ as an extra argument. $EXISTS\_TAC$ maps a goal $\Gamma \stackrel{?}{\vdash} \exists x.A[x]$ to the single goal $\Gamma \stackrel{?}{\vdash} A[t]$. It attacks the goal 'there exists an $x$' by stating 'and that $x$ is in fact $t$.' Unfortunately, the proper choice for $t$, the *existential witness*, may not be evident. A nice analogy has been mentioned: a program runs in order $n^2$ time if it executes $kn^2$ steps, for some constant $k$. A proof that it runs in order $n^2$ time should contain enough information to deduce the value of $k$, but we will never find a proof if we have to guess $k$ as the first step.

## 3  Putting unification into LCF

The LCF tradition favors one-way matching over unification. An inference rule matches theorems of a given pattern; a tactic matches goals. However, some researchers have tried to bring the benefits of unification into LCF.

*Resolution* tactics work on the goal's assumptions, adding new assumptions. Most resolution tactics use one-way matching; Brian Monahan's use unification [28]. Monahan has also automated the construction of simple rules and tactics. His function $METARULE$ turns any theorem $P_1 \wedge \ldots \wedge P_m \Rightarrow Q$ into the rule $\frac{P_1 \cdots P_m}{Q}$. His $METATAC$ produces the corresponding tactic. Their generality is limited because LCF's logic, PP$\lambda$, does not have variables ranging over formulas.

Stefan Sokołowski used Edinburgh LCF to prove the soundness of Hoare axiomatic rules with respect to a denotational semantics of a simple programming language [37]. The proof requires the systematic expansion of many definitions. LCF's simplifier expands definitions by rewriting, but Sokołowski preferred to structure his proof in terms of derived inference rules.

Sokołowski's innovation was to allow *pattern variables* in goals, and allow tactics to instantiate pattern variables by unification [36]. Existential goals are an obvious use for pattern variables. Sokołowski's tactics could allow the existential witness to be inferred later in the proof.[2] An *environment* holds instantiations of pattern variables. A unification tactic takes an environment as well as a goal. It returns an extended environment along with the subgoals and validation. After all subgoals have been solved, the validation is given the final environment as well as a list of theorems.

Sokołowski extended Edinburgh LCF by writing unification tacticals and simple backtracking tactics in ML. There were problems. Edinburgh LCF executed ML slowly. The treatment of environments during sequential composition (the tactical *THEN*) may have been faulty. Each subgoal could instantiate the environment differently; if the environments were incompatible then the proof would fail. I prefer to string a single environment through the subgoals. Even so, Sokołowski's tactics verified the Hoare rules with remarkable clarity, capturing the high-level structure of the proofs.

## 4 Reasoning with inference rules

Schmidt argues that inference rules are more natural than axioms for goal-directed proof [34]. To illustrate the point, he develops natural deduction proof rules from the axioms of Gödel-Bernays set theory. The subset relation $\subseteq$ is defined by the axiom

$$\forall AB.\, A \subseteq B \iff \forall x.\, x \in A \Rightarrow x \in B\,.$$

Reasoning directly from this axiom requires fiddling with many quantifiers and connectives. Schmidt's subset introduction rule is the typical way to prove that $A$ is a subset of $B$:

$$\frac{\Gamma, x \in A \vdash x \in B}{\Gamma \vdash A \subseteq B} \qquad x \text{ not free in } \Gamma$$

His subset elimination rule is the typical use of the knowledge that $A$ is a subset of $B$:

$$\frac{\Gamma \vdash t \in A \qquad \Gamma \vdash A \subseteq B}{\Gamma \vdash t \in B}$$

With these rules we can easily derive new rules, continuing to work at rule level. The proof tree

$$\frac{\dfrac{\dfrac{-}{\Gamma, x \in A \vdash x \in A} \quad \dfrac{\Gamma \vdash A \subseteq B}{\Gamma, x \in A \vdash A \subseteq B}}{\Gamma, x \in A \vdash x \in B} \quad \dfrac{\Gamma \vdash B \subseteq C}{\Gamma, x \in A \vdash B \subseteq C}}{\dfrac{\Gamma, x \in A \vdash x \in C}{\Gamma \vdash A \subseteq C}}$$

---

[2]Sokołowski was using Edinburgh LCF, which lacks existential quantifiers. The same reasoning holds for universally quantified assumptions.

derives the rule
$$\frac{\Gamma \vdash A \subseteq B \quad \Gamma \vdash B \subseteq C}{\Gamma \vdash A \subseteq C}$$

LCF has powerful mechanisms for deriving rules. For forwards proof, an inference rule is a function. A new rule is derived by composing functions. For backwards proof, an inference rule is a tactic. A new tactic is derived from other tactics using *tacticals*, operators designed for this purpose. The tactical *THEN* composes two tactics sequentially. The tactical *REPEAT* composes a tactic with itself repeatedly; it can make a proof tree of arbitrary height, depending on the goal.

Yet LCF's derived rules and tactics do not fully support reasoning about inference rules. In the logic PPλ, variables range over individuals, not formulas or predicates. An induction scheme such as
$$\frac{\vdash A(0) \quad A(n) \vdash A(n+1)}{\vdash A(m)}$$

must be implemented by a complicated function or tactic that takes the formula $A$ as an argument. Applying the function to some formula performs the derivation by executing all the primitive inferences. Execution can be slow, and can fail. The only way to inspect a derived rule or tactic is to test it: a function cannot be printed.

One solution is Gordon's higher-order logic [14]. A formula is simply a term of type *bool*; the theorem
$$\vdash \forall A.\, A(0) \land (\forall n. A(n) \Rightarrow A(n+1)) \Rightarrow \forall m. A(m)$$

expresses the induction scheme. But such use of quantifiers and implication is precisely what Schmidt is trying to escape. So expressed, his subset introduction rule reverts to the axiom it was derived from:
$$\vdash \forall AB.\, (\forall x.\, x \in A \Rightarrow x \in B) \Rightarrow A \subseteq B$$

This approach could be used together with Monahan's *METARULE* and *METATAC*.

Isabelle's solution is to work directly with inference rules, not theorems. Represent inference rules as explicit syntactic structures, not functions. Formal proof consists of composing rules with each other; the resulting proof tree is a derived rule. This unifies the notions of forwards and backwards proof.

Backwards proof takes place by matching a goal with the conclusion of a rule; the premises become the subgoals. Consider the goal $\vdash^{?} (A \land B) \Rightarrow (B \land A)$. Matching it with the conclusion of the ⇒-introduction rule gives the subgoal $A \land B \vdash^{?} B \land A$. Composing this with ∧-introduction gives gives two subgoals: $A \land B \vdash^{?} B$ and $A \land B \vdash^{?} A$. Composing these with the ∧-elimination rules gives two identical subgoals, $A \land B \vdash^{?} A \land B$, which is an instance of the assumption rule. The proof tree is

$$\frac{\dfrac{A \land B \vdash A \land B}{A \land B \vdash B} \quad \dfrac{A \land B \vdash A \land B}{A \land B \vdash A}}{\dfrac{A \land B \vdash B \land A}{\vdash (A \land B) \Rightarrow (B \land A)}}$$

An inference rule is a *scheme*. It stands for the family of inferences obtained by uniformly substituting hypotheses for Γ, formulas for $A$ and $B$, terms for $t$, etc. A rule with no premises is an axiom scheme or theorem scheme. The example above proves the theorem scheme $\vdash (A \wedge B) \Rightarrow (B \wedge A)$. Forwards proof takes place by matching theorem schemes to the premises of a rule, making a new theorem scheme.

Scheme variables may have as decisive an impact on LCF as they did on provers based on Herbrand's theorem. The Davis-Putnam method generated ground clauses. Robinson's resolution allowed variables in clauses, whereby one clause stood for an infinity of ground clauses [6, Chapter 5]. Inference rules are Horn clauses; the composition of rules is resolution. A subtle difference: classical resolution is an *inference rule* for first-order logic; for sentences in clause form, no other inference rule is needed. For Isabelle, resolution builds proof trees from the inference rules of an arbitrary formal logic.

## 5 Formalizing quantifier rules

Quantifiers cause enormous complications. Substitution and bound variables can be handled by representing logical syntax in the typed $\lambda$-calculus. Skolemization can enforce parameter restrictions.

### 5.1 Substitution and the typed $\lambda$-calculus

A expression of the $\lambda$-calculus is a constant, a free variable $\overline{x}$, a bound variable $x$, an abstraction $\lambda x.t$, or a combination $(tu)$. Free variables represent scheme variables in rules, overlined for emphasis.

Types, denoted by Greek letters $\alpha, \beta, \ldots$, are recursively formed. There is a set of *atomic* types, and also *function* types $(\alpha \rightarrow \beta)$. Each expression has a type:

- Each variable and constant has one fixed type.

- If $x$ has type $\alpha$ and $t$ has type $\beta$, then the abstraction $\lambda x.t$ has type $(\alpha \rightarrow \beta)$.

- If $t$ has type $(\alpha \rightarrow \beta)$ and $u$ has type $\alpha$, then the combination $(tu)$ has type $\beta$.

The usual abbreviations save writing needless parentheses. Let $\lambda x_1 x_2 \ldots x_n.t$ abbreviate $\lambda x_1.\lambda x_2.\ldots.\lambda x_n.t$, and $t(u_1, u_2, \ldots, u_q)$ abbreviate $(\cdots((tu_1)u_2)\cdots u_q)$. Let $\alpha_1 \rightarrow \alpha_2 \rightarrow \cdots \rightarrow \beta$ abbreviate $(\alpha_1 \rightarrow (\alpha_2 \rightarrow \cdots \rightarrow (\alpha_p \rightarrow \beta)\cdots))$.

Two $\lambda$-expressions are equal if they can be made identical by a sequence of conversions:

- $\alpha$-conversion is the renaming of a bound variable: $\lambda x.t[x] = \lambda y.t[y]$

- $\beta$-conversion is substitution of an argument into the function body: $(\lambda x.t[x])u = t[u]$

- $\eta$-conversion is extensionality of functions. If $x$ is not free in $t$, and $t$ has function type, then $\lambda x.(tx) = t$

For example, let us represent the syntax of propositional logic. Let *term* be the type of terms, and *form* the type of formulas. Connectives like conjunction and implication are infix constant symbols: let $\wedge$ and $\Rightarrow$ have type $form \rightarrow form \rightarrow form$. A function represents a syntax rule: if $A \in form$ and $B \in form$, then also $A \wedge B \in form$.

All binding operators are represented using $\lambda$. For the universal quantifier, introduce the constant $\Pi$ of type $(term \rightarrow form) \rightarrow form$. Let $B$ be a function from terms to formulas: $B \in term \rightarrow form$. This simply means that $B(t)$ is a formula for all terms $t$. The formula $\Pi(B)$ represents $\forall y.B(y)$. For example, $\lambda x.R(x+0, x)$ has type $term \rightarrow form$.

Applied to a term $t - 3$ it yields the formula $R((t-3)+0, t-3)$, while $\Pi(\lambda x.R(x+0,x))$ represents $\forall x.R(x+0,x)$.

Existential quantifiers can be treated as negated universals, or using another constant $\Sigma$ of type $(term \to form) \to form$. As in Gordon's HOL [13], a theorem prover can parse and print formulas in the usual notation. Quantifiers become $\Pi(B)$ only in the internal representation.

Recall the universal introduction rule:

$$\frac{\Gamma \vdash A[x]}{\Gamma \vdash \forall y.A[y]} \qquad x \text{ not free in } \Gamma$$

Write the quantified formula as $\Pi(B)$. Overline $\overline{\Gamma}$ and $\overline{B}$: they are scheme variables, while the parameter $x$ is not a scheme variable. The rule becomes

$$\frac{\overline{\Gamma} \vdash \overline{B}(x)}{\overline{\Gamma} \vdash \Pi(\overline{B})} \qquad x \text{ not free in } \overline{\Gamma}, \overline{B}$$

The $[\cdots]$ notation, for substitution, has become $(\cdots)$, for function application. There is a further restriction on $x$: it may not appear in $B$ itself. The dependence of $B(x)$ upon $x$ must be purely function application. Otherwise there would be free occurrences of $x$ in the conclusion.

The $\forall$-elimination rule allows a universal theorem to be specialized to a term $t$. Conventional syntax:

$$\frac{\Gamma \vdash \forall x.A[x]}{\Gamma \vdash A[t]}$$

Representation:

$$\frac{\overline{\Gamma} \vdash \Pi(\overline{B})}{\overline{\Gamma} \vdash \overline{B}(\overline{t})}$$

The types *term* and *form* and the connective $\wedge$ are not part of the general framework; even the assertion sign $\vdash$ is just another constant. This representation of syntax is essentially Martin-Löf's *theory of expressions* [10] extended to allow more than one atomic type. Church's $\lambda$-calculus representation of higher-order logic [13] is similar. Church allows quantification over any type, with a different $\Pi$ for each; its formalization in Isabelle is likely to resemble Martin-Löf's $\Pi$. The approach encompasses first-order and higher-order, constructive and classical logics.

## 5.2 Parameters and Skolemization

Parameter restrictions are currently enforced by Skolemization, a representation resulting from (too much) trial and error. The end of this section describes a more natural approach that I intend to try.

The $\forall$-introduction rule, even using the $\Pi$ representation, is not ready to be automated: it contains a parameter restriction. Apart from this restriction, the parameter name has no significance. The rule might as well specify a name that can never appear in $B$ or $\Gamma$ — neither now, nor later by instantiation of $B$ and $\Gamma$.

Extend the object syntax with a family of subscripted parameter symbols $\mathbf{all}_{\Gamma,B}$ for all hypotheses $\Gamma$ and formulas $B$. Now symbols and expressions are mutually recursive: $\Gamma$ and $B$ are part of the symbol $\mathbf{all}_{\Gamma,B}$. Clearly $\mathbf{all}_{\Gamma,B}$ cannot appear in $\Gamma$ or $B$ since expressions are finitely constructed. The occurs check of unification enforces the restriction, 'not free in $\Gamma, B$.' Using the new parameters, $\forall$-introduction becomes a pure Horn clause:

$$\frac{\Gamma \vdash B(\mathbf{all}_{\Gamma,B})}{\Gamma \vdash \Pi(B)}$$

The $\exists$-elimination rule also has parameter restrictions:

$$\frac{\Gamma \vdash \exists y.A[y] \quad \Gamma, A[x] \vdash C}{\Gamma \vdash C} \quad x \text{ not free in } \Gamma, C$$

Representing the quantifier using $\Sigma$ gives

$$\frac{\Gamma \vdash \Sigma(B) \quad \Gamma, B(x) \vdash C}{\Gamma \vdash C} \quad x \text{ not free in } \Gamma, B, C$$

Decorating the parameter **exi** with the expressions it must not occur in gives

$$\frac{\Gamma \vdash \Sigma(B) \quad \Gamma, B(\mathbf{exi}_{\Gamma,B,C}) \vdash C}{\Gamma \vdash C}$$

Schmidt also suggests natural deduction rules using Skolemization [35]. He tags a parameter with free variables rather than with entire expressions.

Perhaps **all** should not include the hypothesis, the subscript $\Gamma$. The simpler and more efficient rule

$$\frac{\Gamma \vdash B(\mathbf{all}_B)}{\Gamma \vdash \Pi(B)}$$

resembles Robinson's [33]. His logic includes *exemplification terms*, a version of Hilbert's $\epsilon$-operator. The rule allows $\mathbf{all}_B$ to appear in the hypotheses $\Gamma$. It is sound with respect to models that assign $\mathbf{all}_B$ a value $y$, if such exists, to make $B(y)$ false. If $B(\mathbf{all}_B)$ is true then so is $\Pi(B)$.

Thus the symbol $\mathbf{all}_{\Gamma,B}$ can be understood syntactically as a way of choosing a name, or semantically as a choice function. An alternative is the function application $\mathbf{all}(\Gamma, B)$, with **all** (now a function symbol) and $\mathbf{all}(\Gamma)$ as subexpressions. In an experiment, such subexpressions grossly expanded the search space in higher-order unification.

Many people object to Skolemization. In a real proof, where $\Gamma$ and $B$ are large expressions, printing $\mathbf{all}_{\Gamma,B}$ requires exponential space on the page. Isabelle has an algorithm for compressing Skolem names to unique identifiers, but this works only for printing, not input. A derived rule, like subset introduction, may contain an unnatural Skolem parameter consisting of the Skolem parameters of its derivation. Unique names can be generated by numbering (Lisp's GENSYM). But this alone does not prevent the parameter from sneaking into $\Gamma$ or $B$ by later instantiation.

Lincoln Wallen tells me that an acyclic graph could enforce the parameter restrictions. The rule

$$\frac{\Gamma \vdash B(x)}{\Gamma \vdash \Pi(B)} \quad x \text{ not free in } \Gamma, B$$

could be represented as a Horn clause together with a tiny graph: the parameter $x$ would point to $\Gamma$ and $B$. Instantiation of variables in $\Gamma$ and $B$ would extend the graph. During resolution of two rules, their graphs would be merged. The occurs check would follow edges leading from parameters, preventing the introduction of cycles. Resolution requires scheme variables to be *standardized apart*: renamed to avoid clashes with variables in the other rule. The parameter names would also be standardized apart, preserving their uniqueness.

Wallen [39] applies his technique to first-order logic and a modal logic. Miller [23] describes a similar technique in the setting of higher-order logic.

# 6 Higher-order unification

Unifying two expressions $t$ and $u$ means solving the syntactic equation $t = u$ by instantiating some of its variables. For ordinary unification, expressions are recursively built up from variables $\overline{x}$ and function applications $F(t_1, \ldots, t_q)$. Two expressions are equal only if they are identical. Expressions and variables are untyped. There are practical algorithms for computing the most general unifier of two expressions, or reporting that no unifier exists.

*Higher-order* unification amounts to solving equations in the typed $\lambda$-calculus with respect to $\alpha$, $\beta$, and possibly $\eta$-conversion [17, 19]. It is *semi-decidable*: if the expressions cannot be unified, the search for unifiers may diverge. Although a complete set of unifiers can be recursively enumerated, it may be infinite. Unifying the expression $\overline{f}(\overline{x})$ with the constant $A$ gives the two unifiers $\{\overline{f} = \lambda y.A\}$ and $\{\overline{f} = \lambda y.y,\ \overline{x} = A\}$. If $A$, $B_1$, ..., $B_q$ are distinct constants, then unifying $\overline{f}(\overline{x})$ with $A(B_1, \ldots, B_q)$ gives $q + 2$ unifiers:

$$\{\overline{f} = \lambda y.A(B_1, \ldots, B_q)\}$$
$$\{\overline{f} = \lambda y.y, \quad \overline{x} = A(B_1, \ldots, B_q)\}$$
$$\{\overline{f} = \lambda y.A(B_1, \ldots, B_{i-1}, y, B_{i+1}, \ldots, B_q), \quad \overline{x} = B_i\} \qquad i = 1, \ldots, q$$

Unifying $\overline{f}(\overline{x}_1, \overline{x}_2)$ with $A(B_1, \ldots, B_q)$ gives $q^2 + q + 3$ unifiers. Too many variables make the search space explode.

For representing first-order logic, second-order expressions suffice: no function variable need have functions as arguments. Marek Zaionc [40] gives a third-order example: the expressions $\overline{f}(\lambda x.x)$ and $A$ have the infinite set of unifiers

$$\{\overline{f} = \lambda y.A\} \qquad \{\overline{f} = \lambda y.y(A)\} \qquad \{\overline{f} = \lambda y.y(y(A))\} \qquad \text{and so on.}$$

## 6.1 Huet's search procedure

Most implementations use Huet's procedures SIMPL and MATCH [17]. SIMPL essentially does first-order unification. Each expression is put into head normal form:

$$\lambda x_1 \ldots x_n. F(t_1, \ldots, t_p) ,$$

where $F$ is a constant, free variable, or bound variable. Such an expression is called *rigid* if $F$ is a constant or a bound variable, and *flexible* if $F$ is a free variable.

Outermost lambdas are stripped off; the bound variables $x_1 \ldots x_n$ become part of the context, behaving much like constants. The input pair of expressions is broken into a set of *disagreement pairs* to be unified. A *rigid-rigid* pair $\langle F(t_1, \ldots, t_q), F(u_1, \ldots, u_q) \rangle$ is simplified to the set of pairs $\langle t_1, u_1 \rangle, \ldots, \langle t_q, u_q \rangle$. If $F \neq G$ then $\langle F(t_1, \ldots, t_p), G(u_1, \ldots, u_q) \rangle$ is recognized as non-unifiable.

A pair $\langle \overline{x}, t \rangle$ has the most-general unifier $\{\overline{x} = t\}$ if $\overline{x}$ does not occur in $t$. MATCH can find this unifier, but it is more efficient if SIMPL instantiates $\overline{x}$ immediately. Function variables complicate the occurs check: $\overline{x}$ and $\overline{f}(\overline{x})$ are unified by both $\{\overline{f} = \lambda y.\overline{x}\}$ and $\{\overline{f} = \lambda y.y\}$. Huet's *rigid path* occurs check gives a practical sufficient condition for $\overline{x}$ and $t$ to be unifiable while detecting some non-unifiable cases. Some cases cannot be easily classified as unifiable or not; the implementor must decide whether an expensive search or a missed unifier is the lesser evil.

MATCH guesses instantiations of variables, usually function variables. A *flex-rigid* pair

$$\langle \overline{f}(t_1, \ldots, t_p), F(u_1, \ldots, u_q) \rangle$$

gives rise to as many as $p+1$ different substitutions for $\overline{f}$. For $i = 1, 2, \ldots$, let $\overline{h}_i(x_1, \ldots, x_p)$ be a new variable of appropriate type. We have

$$\overline{f} = \lambda x_1 \ldots x_p.\, x_i(\overline{h}_1(x_1, \ldots, x_p), \ldots, \overline{h}_m(x_1, \ldots, x_p)) \quad \text{for certain } i,\ \text{by } \textit{projection}$$
$$\overline{f} = \lambda x_1 \ldots x_p.\, F(\overline{h}_1(x_1, \ldots, x_p), \ldots, \overline{h}_q(x_1, \ldots, x_p)) \quad \text{by } \textit{imitation}$$

Imitation applies whenever $F$ is a constant, not a bound variable. Projection applies for those $i$ such that the type of $x_i$ allows $m$ to be chosen to give $\overline{f}$ the correct type. If the type of $\overline{f}$ is $\alpha_1 \to \cdots \to \alpha_p \to \beta$, then $\alpha_i$ must be $\gamma_1 \to \cdots \to \gamma_m \to \beta$. The second-order case is simpler; the projections are

$$\overline{f} = \lambda x_1 \ldots x_p.x_i$$

for those $i$ such that the argument type equals the result type, $\alpha_i = \beta$.

MATCH gives a choice of substitutions, creating an OR tree. One of these substitutions is chosen, applied to all the disagreement pairs, and the search continues. SIMPL immediately rejects a projection if $t_i$ begins with a constant different from $F$; imitation reduces the disagreement pair to the pairs

$$\langle \overline{h}_1(t_1, \ldots, t_p), u_1 \rangle, \quad \ldots, \quad \langle \overline{h}_q(t_1, \ldots, t_p), u_q \rangle\,.$$

Sometimes both expressions begin with a free variable, the *flex-flex* case. Unifying $\overline{f}(t_1, \ldots, t_p)$ with $\overline{g}(u_1, \ldots, u_q)$ yields an explosion of counterintuitive solutions [19]. Huet's algorithm reports success when only flex-flex pairs remain: there is a trivial unifier

$$\{\overline{f} = \lambda x_1 \ldots x_p.\overline{h},\ \ \overline{g} = \lambda y_1 \ldots y_q.\overline{h}\}\,.$$

Since the trivial unifier throws away too much information, a theorem prover should store the flex-flex pairs as constraints on future unifications.

## 6.2 Discussion

Here is an example: unifying $\overline{f}(C, \overline{x})$ with $A(B)$, where $A$, $B$, and $C$ are distinct constants. SIMPL returns $\langle \overline{f}(C, \overline{x}), A(B) \rangle$. MATCH guesses three instantiations for $\overline{f}$:

- $\overline{f} = \lambda yz.y$, by projection : SIMPL finds $\langle C, A(B) \rangle$, a failure node.

- $\overline{f} = \lambda yz.z$, by projection : SIMPL instantiates $\overline{x}$ and returns success, the unifier $\{\overline{f} = \lambda yz.z,\ \overline{x} = A(B)\}$.

- $\overline{f} = \lambda yz.A(\overline{g}(y, z))$, by imitation : SIMPL returns $\langle \overline{g}(C, \overline{x}), B \rangle$. MATCH guesses three instantiations for $\overline{g}$:

  - $\overline{g} = \lambda yz.y$, by projection : SIMPL finds $\langle C, B \rangle$, a failure node.
  - $\overline{g} = \lambda yz.z$, by projection : SIMPL instantiates $\overline{x}$ and returns success, the unifier $\{\overline{f} = \lambda yz.A(z),\ \overline{x} = B\}$
  - $\overline{g} = \lambda yz.B$, by imitation : SIMPL returns success, $\{\overline{f} = \lambda yz.A(B)\}$

This search terminated with three unifiers. For an exercise, work through Zaionc's example. The original disagreement pair soon recurs in the search tree, giving an infinite set of unifiers.

Higher-order unification is effective if we use function variables with care. Most disagreement pairs are simple assignments $\overline{x} = t$; even assignments of function variables are easy. Solving a recursive equation like $\overline{f}(t) = u(\overline{f})$ is hard, often causing the search to diverge. Although the nonrecursive equation $\overline{f}(t_1, \ldots, t_p) = u$ has the trivial solution $\{\overline{f} = \lambda x_1 \ldots x_p.u\}$, finding its interesting solutions requires the full search. There are many different ways of *projecting* onto the arguments $t_1, \ldots, t_p$ or *imitating* parts of $u$.

For example, the equation $\overline{f}(0) = 0 + 0$ has four solutions in $\overline{f}$: $\lambda x.x + x$, $\lambda x.x + 0$, $\lambda x.0 + x$, and $\lambda x.0 + 0$. The first solution is the most natural; the second and third are sometimes useful. The last solution, produced by pure imitation, is only rarely useful. MATCH should try projection before imitation, to produce the trivial solution last.

Are types essential? Logic programming traditionally uses no types. Higher-order unification makes sense for a single atomic type *data* containing integers, booleans, strings, and, recursively, tuples of *data*. But function types must be distinguished from atomic types. In the untyped $\lambda$-calculus, it is undecidable whether an expression has a normal form, so SIMPL could diverge. Worse, there is a fixedpoint combinator $Y$ such that $Y(f) = f(Y(f))$ for all expressions $f$. This solves any disagreement pair in MATCH: $\overline{f}(t) = u(\overline{f})$ has the trivial solution $\{\overline{f} = Y(\lambda fx.u(f))\}$.

Huet gives a version of MATCH for unification without $\eta$-conversion [17]. The search space expands: $f$ need not equal $\lambda x.f(x)$, so many more independent substitutions are possible. These contribute nothing even in a logic with intensional functions, provided that $\lambda$-abstraction is only used as the syntactic representation of binding operators.

Theorem proving is undecidable, but it is unfortunate that each resolution step is undecidable. We can recover decidability by restricting unification. Limiting the search gives unpredictable results. Second-order matching is decidable [18], though second-order unification is not [11]. Perhaps first-order unification plus second-order matching is a practical compromise. Ketonen's EKL proves theorems using first-order unification plus higher-order matching. Ketonen claims that higher-order matching is decidable, without proof [20]; Huet tells me that decidability is an open question.

Quantifiers go beyond first-order unification, but where do we stop? LCF inference rules form the union of hypotheses:

$$\frac{\overline{\Gamma} \vdash \overline{A} \quad \overline{\Delta} \vdash \overline{B}}{\overline{\Gamma}, \overline{\Delta} \vdash \overline{A} \wedge \overline{B}}$$

Must unification must also handle union, an associative, commutative, and idempotent operator? Fortunately, a different treatment of assumptions is possible:

$$\frac{\overline{\Gamma} \vdash \overline{A} \quad \overline{\Gamma} \vdash \overline{B}}{\overline{\Gamma} \vdash \overline{A} \wedge \overline{B}}$$

Unification cannot do everything.

## 7  The implementation

Isabelle consists of 3200 lines of the new Standard ML [26], compiled by David Matthew's Poly/ML on a VAX/750 running Berkeley Unix. Both the language and Matthew's compiler have been assets. Type-checking means that sophisticated code often works first time. Compiled code runs fast: examples with dozens of higher-order unifications run in seconds.

For the typed $\lambda$-calculus, the ML type *arity* represents types and the ML type *term* represents $\lambda$-expressions. Scheme variables are represented by (string,integer) pairs. Each bound variable is also represented by an integer, referring to the depth at which it is bound [4]. The environment primitives use Boyer and Moore's *structure sharing* [3] to standardize variables apart before unification. Normalization, $\alpha$-convertibility, and substitution functions are provided. The $\lambda$-expression parser and printer are extensible. Each can be invoked in mutual recursion with parsing/printing functions written for a particular logical syntax.

Higher-order unification returns a possibly infinite stream of unifiers. Streams are implemented as usual: each member contains a function for computing the rest of the stream. The occurs check is slow: each assignment $\overline{x} = t$ requires scanning $t$ for $\overline{x}$. Omitted in Prolog, the occurs check is essential for enforcing parameter restrictions. Large expressions, representing assumption lists, must be scanned.

The type *rule* and the function *Resolve* provide the basic inference mechanism. In backwards proof, the goal tree is a rule $\frac{Q_1 \cdots Q_n}{P}$. The root $P$ is the initial goal, the leaves $Q_1 \cdots Q_n$ are the unanalyzed subgoals. Resolving a subgoal with a rule produces a new goal tree. At present, Isabelle does not store the tree of resolution steps underlying a rule; as a result, a goal tree has no internal subgoal structure.

An Isabelle tactic is a function on inference rules, regarded as goal trees. It has type *rule* $\rightarrow$ (*rule stream*): it may return a stream of trees. A tactic for goal-directed proof just replaces some leaves, but any function on goal trees is a tactic. *RulesTac* unifies the conclusion of several rules against a goal. There are two levels of choice: several rules may apply; a rule may have several unifiers with the goal. The tactical *DEPTH_FIRST* repeatedly applies a tactic in depth-first search. If the tactic returns an empty stream of goal trees, then the goal is abandoned. Sokołowski used depth-first search for backwards chaining along hypotheses [37]. It is also effective with introduction rules. Together with *RulesTac* it constitutes a higher-order Prolog interpreter. It can execute trivial Prolog programs (slowly).

The interactive goal package maintains the current proof state. When applying a tactic, the package uses the first set of subgoals produced, saving the remainder of the stream. The user can explicitly *backtrack* any past step: discard the current goal tree, taking the next tree from that point. The backtrack command fails if the stream for that step is empty.

In the original goal, scheme variables are useful placeholders for information that is irrelevant or unknown. New variables crop up as existential witnesses. During the proof, scheme variables develop gradually and naturally. A variable may occur in more than one goal; instantiation affects all goals simultaneously.

To minimize backtracking, goals must be tackled in a sensible order. A variable in a critical place leaves the search unconstrained; a goal is *too flexible* if its variables allow too many rules to unify. Experimental depth-first tacticals control the search by expanding only 'appropriate' goals. One argument of *DEPTH_FIRST* is a predicate for classifying a goal as satisfied or unsatisfied. One argument of *DepthRulesFunTac* is a function for analyzing a goal, computing the list of appropriate rules. The function can defer a goal by returning an empty rule list. Deferred goals can be reconsidered as the search proceeds; the search stops when all remaining goals continue to be deferred. A goal can be deferred for any reason, not just if too flexible. A typical application would be to conduct a proof in stages: solving all equational goals, while leaving the others until later.

# 8 Experiments with Constructive Type Theory

Isabelle has been set up for Martin-Löf's Constructive Type Theory, with a parser and printer for its syntax, and tactics for solving typical problems. This thirty percent of the code is kept separate from the rest.

## 8.1 Constructive Type Theory

Martin-Löf's Type Theory is an attempt to formalize constructive reasoning [21, 29]. It interprets *propositions as types*: the rules for each logical connective express its constructive meaning as operations on proof objects, elements of the corresponding type. For instance, the proposition $A \wedge B$ is interpreted as the Cartesian product $A \times B$: a proof of $A \wedge B$ is a pair $\langle a, b \rangle$, where $a$ is a proof of $A$ and $b$ is a proof of $B$. The proof objects form a simple functional programming language. All computations terminate, though the set of functions is much larger than the set of primitive recursive functions.

People are using Constructive Type Theory for program verification and derivation. By 'propositions as types', a small family of primitives provides a full system of logical connectives, data structures, and programs. A type can express a complete formal specification: the type of a sorting function can assert that its output is a sorted permutation of its input. Petersson has implemented Type Theory by modifying Edinburgh LCF [32]. PRL supports a related type theory using ideas from LCF [9].

Type Theory has several kinds of rules:

- *Formation* rules build types from other types.

- *Introduction* rules build elements of types. By 'propositions as types', they also introduce logical connectives.

- *Elimination* rules specify control structures, called *selectors*, for each type: discrimination for sum types, projections for product types, application for function types, and primitive recursion for recursive types. By 'propositions as types', they also eliminate logical connectives.

- *Equality* rules give the result of evaluating expressions.

## 8.2 Functions and the theory of expressions

A higher-order syntax is practically essential. While predicate calculus has only two binding operators, $\forall$ and $\exists$, Type Theory has $\Pi$, $\Sigma$, **split**, **when**, etc., with various binding rules. Martin-Löf's *theory of expressions* [10] is formally equivalent to the typed $\lambda$-calculus with the single ground type (). The notation and terminology are different, especially for types. Martin-Löf uses the word *arity*, reserving *type* for the types described by the inference rules. Let us adopt this convention for now.

There are two sets of notation because there are two kinds of function: the arity $\alpha \rightarrow \beta$ and the (Martin-Löf) type $A \Rightarrow B$. If $A$ and $B$ are types, then $A \Rightarrow B$ is the type of functions from $A$ to $B$, and corresponds to logical implication. Call an expression of arity $\alpha \rightarrow \beta$ a *function*, and an element of type $A \Rightarrow B$ a *function object*. An expression of function arity is also called *unsaturated*, while an expression of atomic arity is called *saturated*. These concepts are due to Gottlob Frege [1]. Functions play important roles in the rules, but only saturated expressions can denote types or elements of types: a single expression cannot be both a function and a function object.

Consider the function $\lambda xy.R(x + y, x)$. Martin-Löf writes this as $(x, y)R(x + y, x)$, reserving $\lambda$ for function objects. I find this hard to read, preferring the traditional $\lambda$-notation, or $\lambda(x, y)R(x + y, x)$ as a compromise. I use **lambda** for abstraction of function objects. If $b$ is a function, then its extension is **lambda**$(x)b(x)$, a function object. There are also two forms of application: for functions, $b(t)$; for function objects, $f * t$ or **apply**$(f, t)$. The Type Theory rule for $\beta$-conversion defines application of function objects in terms of application of functions: $(\textbf{lambda}(x)b(x)) * t = b(t)$. The $\eta$-conversion rule is **lambda**$(x)(c * x) = c$. These rules are needed even though the syntactic theory has both $\beta$ and $\eta$ conversion.

The distinction between functions and function objects is not just for Type Theory. There will be other logics whose notion of function cannot be identified with the $\lambda$-calculus functions. The $\lambda$-calculus representation does not prejudice the notion of function in the logic.

### 8.3 The rules on the computer

The theorems of Type Theory are called *judgements*. A judgement of the form $A$ **type** means that $A$ is a type, while $a \in A$ means that $a$ is an element of type $A$. A judgement can have assumptions, an *ordered* list $x_1 \in A_1$, ..., $x_n \in A_n$. Using an assumption $a \in A$ means searching down the list for $a$, then verifying that $A$ is a type. The Isabelle rules use the Prolog style of list processing:

$$\frac{\overline{\Gamma} \vdash \overline{A} \textbf{ type}}{\overline{\Gamma}, \overline{a} \in \overline{A} \vdash \overline{a} \in \overline{A}} \qquad \frac{\overline{\Gamma} \vdash \overline{a} \in \overline{A}}{\overline{\Gamma}, \overline{b} \in \overline{B} \vdash \overline{a} \in \overline{A}}$$

The product introduction rule is handled like $\forall$-introduction. First of all, Martin-Löf uses the binding operator $\Pi$ for product types. The type $\prod_{y \in A} B(y)$ corresponds to the proposition $\forall y \in A \,.\, B(y)$. Elements of the type are function objects. The rule is written

$$\frac{\Gamma \vdash A \textbf{ type} \quad \Gamma, x \in A \vdash b(x) \in B(x)}{\Gamma \vdash \textbf{lambda}(y)b(y) \in \prod_{y \in A} B(y)} \qquad x \text{ not free in } \Gamma$$

In the $\lambda$-calculus syntactic representation, the binding operators $\Pi$ and **lambda** are used as constant symbols. The representation of $\prod_{y \in A} B(y)$ is $\Pi(A, B)$; the representation of **lambda**$(x)b(x)$ is **lambda**$(b)$. Then making scheme variables explicit gives

$$\frac{\overline{\Gamma} \vdash \overline{A} \textbf{ type} \quad \overline{\Gamma}, x \in \overline{A} \vdash \overline{b}(x) \in \overline{B}(x)}{\overline{\Gamma} \vdash \textbf{lambda}(\overline{b}) \in \Pi(\overline{A}, \overline{B})} \qquad x \text{ not free in } \overline{\Gamma}, \overline{b}, \overline{B}$$

Using the Skolem constant $\textbf{pri}_{\overline{\Gamma},\overline{b},\overline{B}}$, for **pr**oduct **i**ntroduction, gives

$$\frac{\overline{\Gamma} \vdash \overline{A} \textbf{ type} \quad \overline{\Gamma}, \textbf{pri}_{\overline{\Gamma},\overline{b},\overline{B}} \in \overline{A} \vdash \overline{b}(\textbf{pri}_{\overline{\Gamma},\overline{b},\overline{B}}) \in \overline{B}(\textbf{pri}_{\overline{\Gamma},\overline{b},\overline{B}})}{\overline{\Gamma} \vdash \textbf{lambda}(\overline{b}) \in \Pi(\overline{A}, \overline{B})}$$

In Isabelle, the Type Theory parser, *read_rule*, is called from ML. The premises are an ML list of strings; each variable is indicated by a prefixed question mark; Skolem subscripts are in square brackets:

```
val ProdIntrRl = read_rule
( [ "?H |- ?A type",
    "?H, pri[?b1,?B1]: ?A |- ?b1(pri[?b1,?B1]) : ?B1(pri[?b1,?B1])" ] ,
{ ------------------------------------------------------------------ }
    "?H |- lambda(?b1) : Prod(?A,?B1)"  );
```

The horizontal line is simply an ML comment. Omitting the subscript $\overline{\Gamma}$ (ASCII syntax ?H) is an old mistake that I have not gotten round to fixing, since there are a dozen similar rules.

## 8.4 Tactics

Tactics, built from standard primitives, can solve problems expressed in Type Theory. Checking that a type $A$ is well-formed takes place by using formation rules to prove the judgement $A$ **type**.

By unification, a rule can specify more than one direction of information flow. In Prolog this is called *multi-mode* execution: programs can run backwards. The judgement $a \in A$ has several meanings:

- It can mean $a$ is a program of type $A$. Given a program $a$, proving $a \in \overline{A}$ determines the type of $a$. Type inference comes for free, while Gothenburg's Type Theory system [32] directly implements Milner's algorithm, a lot of code using explicit unification [25]. A group at INRIA also obtain the effect of Milner's algorithm by executing inference rules [7].

- It can mean $a$ is a proof of the proposition $A$. Given a proposition $A$, proving $\overline{a} \in A$ gives a *constructive proof* of $A$. While this is undecidable, repeated use of introduction rules performs a large portion of the proof.

- It can mean $a$ is a program with specification $A$. Given a specification $A$, proving $\overline{a} \in A$ amounts to the *program synthesis* of $\overline{a}$.

The tactic *DepthIntrTac* uses *DepthRulesFunTac* to handle variables sensibly in the formation and introduction rules. The goal $A$ **type** is deferred if $A$ is a variable. For $a \in A$ either $a$ or $A$ must not be a variable. *TypeCheckTac* is similar but uses elimination rules as well as formation and introduction rules. For $a \in A$ it requires $a$ to be rigid, since if $a$ is a variable then all the elimination rules apply. *TypeCheckTac* handles the type-checking problems that have come up in my experiments. It can check the type of the addition operator, as defined by primitive recursion. Solving the goal

$$\textbf{lambda}(k)\,\textbf{lambda}(m)\,\textbf{rec}(m,\,\lambda(x,y)\,\textbf{succ}(y),\,k) \in \overline{A}$$

assigns $\overline{A} = \textbf{Nat} \Rightarrow \textbf{Nat} \Rightarrow \textbf{Nat}$.

A Type Theory function takes apart its arguments using the awkward selector operators. An ML function is defined by equations on its patterns of input. To support a pattern-directed style for Type Theory, I have experimented with tactics for manipulating equations. Some examples go through with little guidance. For the *predecessor function* the tactics discover that $pred = \textbf{lambda}(x)\,\textbf{rec}(x,\lambda(y,z)y,x)$ with type $\overline{A} = \textbf{Nat} \Rightarrow \textbf{Nat}$ by solving the goal

$$\overline{x} \in \sum_{pred \in \overline{A}} \left( pred * 0 = 0 \,\times\, \prod_{k \in \textbf{Nat}} pred * \textbf{succ}(k) = k \right).$$

The function *fst*, on product types, takes the first component of a pair $\langle a, b \rangle$. The tactics discover that $fst = \textbf{lambda}(x)\,\textbf{split}(\lambda(y,z)y,x)$ with type $\overline{A} = (\textbf{Nat} \times \textbf{Nat}) \Rightarrow \textbf{Nat}$:

$$\overline{x} \in \sum_{fst \in \overline{A}} \prod_{i \in \textbf{Nat}} \prod_{j \in \textbf{Nat}} fst * \langle i, j \rangle = i$$

This example involves both product and sum types. The tactics discover a binding for $\overline{f}$ using the selectors **when** and **split**.

$$\overline{x} \in \prod_{i \in \mathbf{Nat}} \prod_{j \in \mathbf{Nat}} \overline{f}(\mathbf{inl}\langle i,j \rangle) = i \ \times \ \overline{f}(\mathbf{inr}\langle i,j \rangle) = j$$

Such use of function variables can cause unification problems. An arithmetic addition *function* can be discovered, but addition as a *function object* of type $\mathbf{Nat} \Rightarrow \mathbf{Nat} \Rightarrow \mathbf{Nat}$ performs computation on both numbers and function types. Its derivation produces subgoals containing unsimplified recursion equations; unification diverges.

## 9 Related work

The earliest applications of higher-order unification extended resolution to higher-order logic [19]. Huet's *constrained resolution* postponed branching in unification [16]. Rather than returning multiple unifiers in a resolution step, it recorded the remaining disagreement pairs as constraints on the new clause. Further resolutions satisfied the constraints or rendered them clearly non-unifiable. Constrained resolution went beyond using only flex-flex disagreement pairs, which are always unifiable, as constraints.

The TPS theorem prover uses sophisticated heuristics in the search for higher-order unifiers [24]. In MATCH it chooses a disagreement pair likely to cause the least branching of the tree. It hashes disagreement sets to determine whether a new set is subsumed by an older one. Though the subsumption test is expensive it cuts the search space substantially and prevents some searches from diverging. TPS uses *general matings* rather than resolution. The mating approach unifies subformulas against each other without reducing everything to clause form. TPS can automatically prove Cantor's Theorem: every set has more subsets than elements [2]. Unification discovers the diagonalization function.

The EKL proof checker uses higher-order matching of rewrite rules [20]. N. G. de Bruijn's AUTOMATH project has investigated several higher-order $\lambda$-calculi, reminiscent of Martin-Löf's type theory, as languages for machine-checked proof [5]. Huet and Coquand's *theory of constructions* is a natural development from AUTOMATH [8]. Gordon's HOL is a version of LCF for proving theorems in Church's higher-order logic [13]. The logics of HOL, EKL, and TPS are all descended from Church's.

Gilles Kahn and his group execute operational semantics expressed as inference rules. The rules are preprocessed, then translated into Prolog. They have considered the dynamic semantics of several simple languages, and ML type-checking [7]. The system runs inside the structure editor *Mentor*, providing type-checking and execution of the program being edited.

## 10 Future

Though Isabelle can handle small examples, much work remains before interesting proofs can be attempted.

*Quantifiers* have been the number one trouble spot. A list of failed approaches would fill the page. Skolem constants work but are clumsy. I hope that the acyclic dependency graph will work.

LCF's *simplifier* uses equations as rewrite rules. Proving $(i + j) + k = i + (j + k)$ should be trivial: use induction on $i$, then simplify the base and step subgoals. Without a simplifier, this is impossibly tedious. LCF ideas may need drastic change because of unification.

*Higher-order unification* behaves well if not provoked by unreasonable use of function variables. Most of the time first-order unification takes place, so Isabelle does not require the TPS subsumption test. Decidable restrictions of unification should be found.

An LCF *theory* is a data base of constants, types, axioms, and theorems. At present, Isabelle allows only simple abbreviations. We need methods for combining theories and working in different logics.

The *user interface* is crude. In LCF, the logic is integrated with ML; in Isabelle, the parser and printer must be invoked. Goals are designated by number; a high-resolution display and mouse would help.

*Other logics* must be considered to test whether Isabelle is really general. One candidate is the Logical Theory [10], a first-order intuitionistic logic in which Type Theory can be constructed. Something radically different, like a Hoare logic or a temporal logic, should be attempted.

### Acknowledgements

David Matthews has worked hard on his Standard ML compiler, with funding from the Science and Engineering Research Council. The main ideas came from the work of Gérard Huet, Per Martin-Löf, and Stefan Sokołowski. Lincoln Wallen suggested a promising alternative to Skolemization. Gilles Kahn's group demonstrated related work. Thanks also to Peter Aczel, Peter Andrews, Michael Gordon, Dale Miller, Bengt Nordström, Kent Petersson, Frank Pfenning, Alan Robinson, Jan M. Smith, and Richard Waldinger.

# References

[1] P. Aczel, Frege's formal language, Printed notes, Dept. of Mathematics, University of Manchester, England (1981).

[2] P. B. Andrews, D. A. Miller, E. L. Cohen, F. Pfenning, Automating higher-order logic, in: W. W. Bledsoe and D. W. Loveland, editors, *Automated Theorem Proving: After 25 Years*, American Mathematical Society (1984), pages 169–192.

[3] R. S. Boyer, J S. Moore, The sharing of structure in theorem-proving programs, in: B. Meltzer and D. Michie, editors, *Machine Intelligence 7* (Edinburgh University Press, 1972), pages 101–116.

[4] N. G. de Bruijn, Lambda calculus notation with nameless dummies, a tool for automatic formula manipulation, with application to the Church-Rosser Theorem, *Indagationes Mathematicae* **34** (1972), pages 381–392.

[5] N. G. de Bruijn, A survey of the project AUTOMATH, in: J. P. Seldin, J. R. Hindley, *To H. B. Curry: Essays in Combinatory Logic, Lambda Calculus and Formalism* (Academic Press, 1980), pages 579–606.

[6] C.-L. Chang, R. C.-T. Lee, *Symbolic Logic and Mechanical Theorem Proving* (Academic Press, 1973).

[7] D. Clément, J. Despeyroux, T. Despeyroux, L. Hascoet, G. Kahn, Natural semantics on the computer, INRIA Research Report 416, Sophia-Antipolis, France (1985).

[8] T. Coquand, G. Huet, Constructions: a higher order proof system for mechanizing mathematics, in: B. Buchberger, editor, *EUROCAL '85: European Conference on Computer Algebra*, Volume 1: *Invited lectures*, Springer LNCS 203 (1985), pages 151–184.

[9] R. L. Constable, T. B. Knoblock, J. L. Bates, Writing programs that construct proofs, *Journal of Automated Reasoning* **1** (1985), 285–326.

[10] P. Dybjer, Program verification in a logical theory of constructions, in: J.-P. Jouannaud, editor, *Functional Programming Languages and Computer Architecture* (Springer LNCS 201, 1985), pages 334–349.

[11] W. D. Goldfarb, The undecidability of the second-order unification problem, *Theoretical Computer Science* **13** (1981), pages 225–230.

[12] M. J. C. Gordon, Representing a logic in the LCF metalanguage, in: D. Néel, editor, *Tools and Notions for Program Construction*, Cambridge University Press, pages 163–185, 1982.

[13] M. J. C. Gordon, HOL: A machine oriented formulation of higher order logic, Report 68, Computer Lab., University of Cambridge (1985).

[14] M. J. C. Gordon, Why higher-order logic is a good formalism for specifying and verifying hardware, Report 77, Computer Lab., University of Cambridge (1985).

[15] M. J. C. Gordon, R. Milner, and C. P. Wadsworth, *Edinburgh LCF: A Mechanised Logic of Computation*, Springer LNCS 78 (1979).

[16] G. P. Huet, A mechanization of type theory, *Third International Joint Conference on Artificial Intelligence* (1973).

[17] G. P. Huet, A unification algorithm for typed $\lambda$-calculus, *Theoretical Computer Science* **1** (1975), pages 27–57.

[18] G. P. Huet, B. Lang, Proving and applying program transformations expressed with second-order patterns, *Acta Informatica* **11** (1978), pages 31–55.

[19] D. C. Jensen, T. Pietrzykowski, Mechanizing $\omega$-order type theory through unification, *Theoretical Computer Science* **3** (1976), pages 123–171.

[20] J. Ketonen, EKL— A mathematically oriented proof checker, in: R. E. Shostak, editor, *Seventh Conference on Automated Deduction*, Springer LNCS 170 (1984), pages 65–79.

[21] P. Martin-Löf, *Intuitionistic type theory* (Bibliopolis, 1984).

[22] P. Martin-Löf, On the meanings of the logical constants and the justifications of the logical laws, Printed notes, Department of Mathematics, University of Stockholm (1984).

[23] D. A. Miller, Expansion tree proofs and their conversion to natural deduction proofs, in: R. E. Shostak, editor, *Seventh Conference on Automated Deduction*, Springer LNCS 170 (1984), pages 375–393.

[24] D. A. Miller, E. L. Cohen, P. B. Andrews, A look at TPS, in: D. W. Loveland, editor, *Sixth Conference on Automated Deduction*, Springer LNCS 138 (1982), pages 50–69.

[25] R. Milner, A theory of type polymorphism in programming, *Journal of Computer and System Sciences* **17** (1978), pages 348–375.

[26] R. Milner, A proposal for Standard ML, *ACM Symposium on Lisp and Functional Programming* (1984), pages 184–197.

[27] R. Milner, The use of machines to assist in rigorous proof, *Philosophical Transactions of the Royal Society of London* **312** (1984), pages 411-422. Also in: C. A. R. Hoare, J. C. Sheperdson, editors, *Mathematical Logic and Programming Languages* (Prentice-Hall, 1984).

[28] B. Q. Monahan, *Data Type Proofs using Edinburgh LCF*, PhD Thesis, University of Edinburgh (1984).

[29] B. Nordström and J. M. Smith, Propositions and specifications of programs in Martin-Löf's type theory, *BIT* **24** (1984), pages 288–301.

[30] L. C. Paulson, Lessons learned from LCF: a survey of natural deduction proofs, *Computer Journal* **28** (1985), pages 474–479.

[31] L. C. Paulson, Interactive theorem proving with Cambridge LCF: A user's manual, Report 80, Computer Lab., University of Cambridge (1985).

[32] K. Petersson, A programming system for type theory, Report 21, Department of Computer Sciences, Chalmers University, Göteborg, Sweden (1982).

[33] J. A. Robinson, *Logic: Form and Function* (Edinburgh University Press, 1979).

[34] D. Schmidt, Natural deduction theorem proving in set theory, Report CSR-142-83, Dept. of Computer Science, University of Edinburgh (1983).

[35] D. Schmidt, A programming notation for tactical reasoning, in: R. E. Shostak, editor, *Seventh Conference on Automated Deduction*, Springer LNCS 170 (1984), pages 445–459.

[36] S. Sokołowski, A note on tactics in LCF, Report CSR-140-83, Dept. of Computer Science, University of Edinburgh (1983).

[37] S. Sokołowski, Soundness of Hoare's logic: an automatic proof using LCF, *ACM Transactions on Programming Languages and Systems* **9** (1987), pages 100–120.

[38] N. Tennant, *Natural Logic* (Edinburgh University Press, 1978).

[39] L. A. Wallen, Generating connection calculi from tableaux and sequent based proof systems, Research paper 258, Department of Artificial Intelligence, University of Edinburgh (1985).

[40] M. Zaionc, The set of unifiers in typed λ-calculus as regular expressions, *Rewriting Techniques and Applications*, Dijon, France (1985).